\documentclass[12pt]{article}
\usepackage{amsfonts,amsmath}
\usepackage{amssymb}

\renewcommand{\theequation}{\thesection.\arabic{equation}}
\renewcommand{\thesubsection}{\arabic{section}.\arabic{subsection}}
\makeatletter \@addtoreset{equation}{section} \makeatother

{\vspace{3mm} }

\def\al{\alpha}

\def\*{\star}
\def\e{\mathbf{e}}
\def\E2{\mathbf{E}}
\def\w{\mathbf{w}}
\def\R{\mathbf{R}}

\newcommand{\be}{\begin{equation}}
\newcommand{\ee}{\end{equation}}
\newcommand{\bee}{\begin{eqnarray}}
\newcommand{\beee}{\begin{array}}
\newcommand{\eee}{\end{eqnarray}}
\newcommand{\eeee}{\end{array}}
\newcommand{\gb}{\beta}
\newcommand{\gga}{\gamma}

\newcommand{\D}{{\cal D}}

\newcommand{\gd}{\delta}
\newcommand{\gl}{\lambda}

\newcommand{\gep}{\epsilon}

\newcommand{\gs}{\sigma}

\newcommand{\nn}{\nonumber}

\newcommand{\ff}{\frac}

\begin{document}

\begin{flushright}
FIAN/TD/2011-16\\
\end{flushright}

\vspace{0.5cm}

\begin{center}
{\large\bf Coordinate independent approach to $5d$ black holes}

\vspace{1 cm}

{\bf V.E.~Didenko}\\
\vspace{0.5 cm} {\it
I.E. Tamm Department of Theoretical Physics, Lebedev Physical Institute,\\
Leninsky prospect 53, 119991, Moscow, Russia }

\vspace{0.6 cm}
didenko@lpi.ru \\
\end{center}

\vspace{0.4 cm}

\begin{abstract}
\noindent Five-dimensional generalization of $(A)dS_5$-Kerr black
hole is shown to be generated in a coordinate free way by a single
$AdS_5$ global symmetry parameter. Its mass and angular momenta
are associated with Casimir invariants of the background
space-time symmetry parameter leading to the black hole
classification scheme similar to that of relativistic fields
resulting apart from ordinary black hole to ``tachyonic'' and
``light-like'' ones.
\end{abstract}

\section{Introduction}

The generalization of Myers-Perry black holes \cite{MP} to include
a non-zero cosmological constant found in \cite{Gibon} has
provided a great deal of interest in higher-dimensional black
holes. There were several attempts since then to generalize the
result of \cite{Gibon} to add NUT-charge  if not electro-magnetic
charges to the metric altogether. Little progress in the
generalization of higher-dimensional Einstein-Maxwell black holes
has been still achieved. The authors of \cite{CLP} were able,
however, to add the NUT-charge yielding Kerr-NUT-(A)dS metric --
the most general higher-dimensional Einstein black hole solution
available to the moment. This solution in many respects resembles
its four dimensional counterpart represented by
Carter-Pleba\'{n}ski metric \cite{Carter, Pleb}. Particularly,
many ``mysteries'' attributed to $4d$ black holes such as complete
integrability of geodesic equations and variable separation for
Hamilton-Jacobi, Klein-Gordon and Dirac equations in the black
hole background turned out to be resided in higher-dimensional
generalization as well \cite{d-geo, d-KG, d-Dirac}. The origin of
these mysteries in four dimensions was clarified by Floyd and
Penrose \cite{FP} as they found a Killing-Yano tensor \cite{Yano}
responsible for all these miracles put together. In
higher-dimensions a single Killing-Yano tensor is insufficient to
provide complete integrability of the aforementioned equations. It
has been shown recently in \cite{Frolov} that the class of
Kerr-NUT-(A)dS metrics admits the so called principal conformal
Killing-Yano tensor (CYK) that generates a tower of  Killing-Yano
tensors necessary for integrability and variable separation. Of
this CYK field the authors \cite{Frolov} refered to as of
higher-dimensional black hole hidden symmetry.

In the present paper we wish to lay the views on black holes based
on unfolded formulation of dynamical systems \cite{Ann, more}
which being coordinate independent in principal allows us to
identify structures that may be hidden in a particular coordinate
system. For example, the unfolded approach applied to four
dimensional black holes in \cite{DMV} happened to be very
efficient demonstrating that the $(A)dS_4$ black hole is generated
by a single $(A)dS_4$ global symmetry parameter encoding the black
hole mass and angular momentum in its two Casimir invariants. Our
primary interest is higher-spin generalization of general
relativity black holes. This is where the unfolded approach
reveals its strength. It was shown in \cite{DMV0,DMV} that apart
from $4d$ black hole solution in gravity sector, the $(A)dS_4$
global symmetry parameter generates a tower of Kerr-Schild type
solutions of spin $s$ Fronsdal equations. That Kerr-Schild field
{\it i.e.,} shear free geodesic null congruence, generates
solutions of massless field equations has been known ever since
Penrose transform was invented \cite{Penrose}. The main result of
\cite{DMV0,DMV}, however, is that it is the $AdS_4$ global
symmetry parameter that determines the black hole null geodesic
congruence. This is what really important for higher-spin
generalization. To obtain a four dimensional higher-spin black
hole, the global higher-spin symmetry parameter rather than the
$AdS_4$ one was chosen in \cite{DV}. Via an analogue of Penrose
transform it has produced then a solution \cite{DV} of the $4d$
nonlinear higher-spin equations \cite{Vas4, Vas4-more} which boils
down to $AdS_4$-Schwarzschild black hole in gravity sector in the
weak field limit when higher-spin fields decouple.

We would like to stress that despite often quoted, the notion of
higher-spin black hole is yet to be justified. Space-time geometry
in higher-spin field theories is far from being understood. Hence,
the role of the metric in such theories is rather hazy not to
mention that the line element associated with it is generically no
longer gauge invariant \cite{DV}\footnote{It should be noted that
in some cases it is possible to construct Lorentz invariant line
element in higher-spin gravity as has been shown for $sl(3)\oplus
sl(3)$ Chern-Simons theory in \cite{Comp}}. Thus, an event horizon
attributed to black holes as we know them is not well defined in
higher-spin gauge theories. To the moment, we prefer to take a
higher-spin black hole as a solution of higher-spin field
equations that has similar space-time symmetry as ordinary black
hole has. Examples of such solutions can be found in \cite{DV, IS,
Kraus1, Kraus2, Kraus3}. An interesting conjecture on what
actually should be called a higher-spin black hole in three
dimensions is given in \cite{Kraus1} and elaborated further on in
\cite{Kraus2, Kraus3}. The authors of \cite{Kraus1} have found a
static solution of $sl(3)\oplus sl(3)$ Chern-Simons gravity that
carries spin two and spin three charges and conjectured that the
eigenvalues of Chern-Simons black hole connection holonomies
around the Euclidian time circle should be equal to those of the
BTZ black hole \cite{BTZ}. Interestingly, this proposal turns out
to be consistent with the integrability condition coming out from
the thermodynamical partition function. Another curious result on
black holes in higher-derivative theories was obtained in
\cite{derivat}, which shows that higher-derivative terms may
smooth out curvature singularities.

Aiming higher-spin generalization of black holes in $d>4$ we would
like to understand to what extent the results of the four
dimensional analysis carried out in \cite{DMV} can be extended
beyond $d=4$. Particularly, it is of interest to see if the
Kerr-Schild ansatz works for higher-spin massless fields in $d>4$.
On one hand, the generalization of Kerr-Schild fields for higher
spins does not seem feasible as the relation between null shear
free geodesic congruence and massless equations is essentially
four dimensional in nature being based on the Penrose twistor
transform\footnote{In higher-spin theory an analogue of Penrose
transform relates the adjoint module to the twisted-adjoint one in
any dimension being identical to that of Penrose in $d=4$ (see
\cite{GV} for application in $4d$).}. On the other hand, both
$d$-dimensional Myers-Perry example \cite{MP} and its recent
generalization for a nonzero cosmological constant \cite{Gibon}
illustrate that Kerr-Schild anzatz, for some reason, remains valid
for gravity fields in arbitrary $d$. The other observation in
favor of higher-dimensional generalization of the $4d$ description
proposed in \cite{DMV}  is black hole ``hair'' counting. Indeed,
one of the reason why a $4d$ black hole could be generated by an
AdS/Minkowski symmetry parameter is that the rank of the
corresponding isometry algebra $o(3,2)$ or $iso(3,1)$ is equal to
the number of black hole parameters -- mass and angular momentum,
{\it i.e.} 2. In higher dimensions, there are $[\ff{d-1}{2}]$
angular momenta and one mass parameter, altogether
$[\ff{d+1}{2}]=$rank $(o(d-1,2),\,o(d,1),\,iso(d-1,1))$.

The purpose of this paper is to analyze if a general relativity
$5d$ black hole can be generated by background space-time
($(A)dS_5$ or Minkowski) global symmetry parameter. We answer to
this question in the affirmative and provide a coordinate free
black hole classification in the spirit of relativistic fields
classification. There are three Casimir invariants associated with
the symmetry parameter -- one is $P^2$, the other two $I_1, I_2$
are associated with spins. These three identify the resulting
black hole solution in the following way. For $P^2<0$ we reproduce
ordinary Myers-Perry type $5d$ black hole originally obtained in
\cite{HHTR} which we refer to as of the Kerr black hole case. Its
two angular momenta are associated with the spin Casimir
invariants $I_1, I_2$. For $P^2=0$ the solution is called
``light-like'' Kerr since in Newton's limit the ``source'' is
located on a light-like surface rather than on a time-like one as
in the case of ordinary Kerr\footnote{Let us stress that the
notion of source is considered here in Newton's limit and should
not be confused with the true singularity of a black hole metric
which, as well known from the Penrose diagram, belongs to the
space-like world-line.}. For $P^2>0$ the metric corresponds to
``tachyonic'' Kerr. Arbitrary $P^2$ can be identified with the
$5d$ analogue of the Carter-Pleba\'{n}ski parameter $\gep$. Just
as in the four dimensional case, it can be set to be discrete
$P^2=-1,0,+1$. Then, we will show how the background global
symmetry condition can be deformed in such a way that its
integrability condition $d^2=0$ gets consistent for the black hole
curvature tensor. The deformation parameter is associated with the
black hole mass. The resulting system has the unfolded form
analogous to that obtained in \cite{DMV} and can be reduced to the
initial nondeformed one by local field redefinition.

A great deal of computational simplification in classical $4d$
general relativity results from an extensive use of two-component
Weyl spinors. Particularly, Petrov classification of the Weyl
tensor looks especially simple in terms of spinors. In the generic
$d$-dimensional space-time spinor-to-vector isomorphism is hardly
applicable because of the exponential growth of Clifford algebra
dimension compared to vector's linear growth. $d=5$ is still small
enough to take advantage of the spinor approach. Particularly, the
$5d$ analogue of Petrov classification for the Weyl tensor has
been recently established \cite{DeSmet}.\footnote{As shown in
\cite{5dspinors}, De Smet classification \cite{DeSmet} in fact
does not account for the reality condition. Hence, some of the
cases were not possible. In \cite{5dspinors}, this gap has been
filled.} In deriving our results we use $5d$ spinor formalism.

The paper is organized as follows. In section \ref{General} we
give the notation we use throughout the paper, and then recall
some generalities on the Kerr-Schild anzatz for solving Einstein
equations and introduce the global symmetry condition on $(A)dS$
and Minkowski space-time. In section \ref{Spinor}, the spinor form
of this condition is studied, Casimir invariants are constructed.
In subsection \ref{KS-massless}, we build Kerr-Schild vectors and
associate massless fields on $(A)dS_5$ background. In section
\ref{BH}, we generate black hole solutions and classify those on
Minkowski background according to the values of Poincar\'{e}
invariants. In subsection \ref{BHUS}, the unfolded form of the
black hole system as a deformation of the background global
symmetry condition is given. Its properties and its relation to
nondeformed equations are elaborated. Conclusion is given in
section \ref{conc}. To be self-contained, we provide four
appendices. Two of them devoted to Cartan formalism and
five-dimensional spinors, while in the remaining two the proof of
the geodesity condition is given and some useful identities are
summarized.

\section{Generalities}\label{General}
\subsection{Notation}
In this paper, the following notation has been adopted. Latin
indices from the middle of alphabet $m,n,\dots $ are attributed to
space-time tensors and range $0,\dots, d-1$. Latin indices from
the beginning of alphabet $a,b,\dots$ are fiber ones and range
$a,b=0,\dots, d-1$, raised and lowered using mostly plus flat
Minkowski metric $\eta_{ab}=diag(-1,1,\dots, 1)$. The background
covariant derivative is denoted by $D$, while $\D$ is the black
hole one. Finally, $5d$ spinor indices are Greek $\al,\gb,\dots$
and range four values $1,\dots, 4$.

\subsection{Kerr-Schild metric}\label{K-S}
It is well known since \cite{MP} that $d$-dimensional Einstein
black holes with spherical horizon topology admit Kerr-Schild
representation even in the presence of a cosmological constant
\cite{Gibon}. This means that the black hole metric being an exact
solution of Einstein equations has a perturbative form around the
flat or $(A)dS_d$ background
\be\label{KSvec}
g_{mn}=\bar{g}_{mn}+\ff{2M}{H} k_{m}k_{n}\,,\quad
g^{mn}=\bar{g}^{mn}-\ff{2M}{H}k^m k^n\,,
\ee
where $\bar{g}_{mn}$ is the fiducial metric, $M$ is the black hole
mass, $H$ is some function. The fluctuational part has a specific
factorized form with the Kerr-Schild vector $k^m$ being null and
geodesic
\be\label{KS-vec}
k^mk_m=0\,,\quad k^m\D_m k_n=0\,,
\ee
where $\D$ is a covariant with respect to the $g_{mn}$ derivative.
Indices in \eqref{KSvec} and \eqref{KS-vec} can be raised and
lowered by either background or full metric since $k^m$ is
light-like. This makes $k^m$ a vector with respect to both
metrics. Moreover, it can be verified to be geodesic for both
metrics as well
\be
k^m\D_m k_n=k^mD_m k_n=0\,.
\ee
The detailed analysis of $d$-dimensional Kerr-Schild ansatz can be
found in \cite{Pravda}. The key property of the Kerr-Schild
construction is that it renders Einstein equations linear. In
other words, if $\bar{g}_{mn}$ is the $(A)dS_d$ metric such that
its Ricci tensor $\bar{R}_{mn}(\bar{g})=(d-1)\Lambda\bar{g}_{mn}$,
then Einstein equations for \eqref{KSvec} $R_{mn}(g)=(d-1)\Lambda
g_{mn}$ reduce to the first order background free field equations
\be\label{einst}
\Box
h_{mn}-D_{p}D_{m}h^{p}{}_{n}-D_{p}D_{n}h^{p}{}_{m}=-2(d-1)\Lambda
h_{mn}\,,\quad h_{mn}=\ff1Hk_mk_n
\ee
The nonlinear $O(M^2)$  part is satisfied on account of
\eqref{KS-vec} and \eqref{einst} leaving no new constraints for
$k^m$ or $H$.

Finally, the last comment on Kerr-Schild is in order. It was
already mentioned  that $k^m$ is a Kerr-Schild vector for both
metrics. For Myers-Perry black holes, the function $H$ is a scalar
with respect to both metrics. Altogether this implies that the
metric \eqref{KSvec} is given in a background covariant form. As
will be shown, a sufficient ingredient which generates on--shell
metrics in the form \eqref{KS} is a background space-time global
symmetry parameter.

\subsection{Background symmetries}
To make use of the background space-time global symmetry parameter
in the black hole description, it is convenient to work in the
Cartan formalism (see appendix \ref{App-A}). Let
$\w_{ab}=-\w_{ba}=w_{ab,n}dx^n$ be the Lorentz connection 1-form
and $\e_a=e_{a,n}dx^n$ vielbein 1-form. $(A)dS_d$ is encoded in
the following structure equations
\be\label{curv}
d\w_{ab}+\w_{a}{}^{c}\wedge\w_{cb}=\Lambda\,\e_a\wedge \e_b\,,
\ee
\be\label{tors}
D\e_a=d\e_a+\w_{a}{}^{b}\wedge\e_b=0\,.
\ee
The corresponding curvature tensor
$R_{ab,cd}=\Lambda(\eta_{ac}\eta_{bd}-\eta_{bc}\eta_{ad})$ is of
$dS_d$ for $\Lambda>0$,  $AdS_d$ if $\Lambda<0$ or Minkowski for
$\Lambda=0$. Equations \eqref{curv} and \eqref{tors} have manifest
local gauge symmetry
\be\label{gauge}
\gd\w_{ab}=D\xi_{ab}+\Lambda(\xi_a\e_b-\xi_b\e_a)\,,\quad
\gd\e_a=D\xi_a-\xi_{ab}\e^b\,,
\ee
where $\xi_{ab}=-\xi_{ba}$ and $\xi_a$ are arbitrary 0-forms. Any
particular solution of \eqref{curv}, \eqref{tors} breaks down its
local symmetry. The leftover global symmetry is determined by
$\gd\w_{ab}=\gd\e_a=0$, equivalently
\begin{eqnarray}
D\xi_a &=& \xi_{ab}\e^b\,,\label{Kil}\\
D\xi_{ab} &=& -\Lambda(\xi_a\e_b-\xi_b\e_a)\,.\label{CYK}
\end{eqnarray}
The first equation \eqref{Kil} says that $D_a\xi_b$ does not
contain its symmetric part. In other words $\xi_a$ is a Killing
vector
\be
D_a\xi_b+D_b\xi_a=0\,.
\ee
Equation \eqref{CYK} is simply the $(A)dS_d$ consistency condition
arising from $D^2\xi_a=\Lambda\,\e_a\wedge\e_b\,\xi^b$.

Let us note that equation \eqref{CYK} alone (equation \eqref{Kil}
is not imposed) being written in an arbitrary space-time imposes
severe restrictions on its geometry. Those spaces which admit
nontrivial \eqref{CYK} compatible with Bianchi identities possess
hidden symmetries \cite{Frolov}. These symmetries arise from
Yano-Killing tensors associated with $\xi_{ab}$. The field
$\xi_{ab}$ itself was called in \cite{Frolov} the principal
conformal Yano-Killing tensor (CYK). It plays a crucial role in
the variable separation problem for higher-dimensional black
holes. Moreover, in \cite{Frolov2} $2d$-dimensional
Chen-L\"{u}-Pope black holes \cite{CLP} were shown to be the only
on-shell solutions which admit CYK field. As the mass and
NUT-parameter of a black hole are set to zero, the metric reduces
to $(A)dS_d$.

Consider now the flat case with $\Lambda=0$. It will be convenient
to have the same form of equation \eqref{CYK} in this limit.
However, as $\Lambda\to 0$ one arrives at $D\xi_{ab}=0$. A field
redefinition $(\xi_a, \xi_{ab})\leftrightarrow(v_a, \Phi_{ab})$
implies
\be\label{Mink}
Dv_a=0\,,\quad D\Phi_{ab}=v_a\e_b-v_b\e_a\,.
\ee
Equations in \eqref{Mink} are obviously consistent with $D^2=0$
and have the same amount of fields as in \eqref{Kil}, \eqref{CYK}.
This redefinition can be made explicit in the Cartesian reference
frame, for example,  with $\w_{ab}=0$, $\e_a=dx_a$:
\be
v_a=\xi_a-\xi_{ab}x^b\,,\quad
\Phi_{ab}=\xi_{ab}+(\xi_a-\xi_{ac}x^c)x_b
-(\xi_b-\xi_{bc}x^c)x_a\,.
\ee
It is clear that \eqref{Mink} is still a covariant constancy
condition for the Poincar\'{e} case.  Note, that $v^a$ is a
Minkowski Killing vector. Not all Killing vectors satisfy $Dv_a=0$
in Minkowski space-time though the rest - which do not - are
encoded in $\Phi_{ab}$. Using these new fields one rewrites the
global symmetry parameter equation for $(A)dS_d$ or Minkowski
space-time in a uniform manner that preserves the CYK
equation\footnote{Factors of two and one-half in
\eqref{main1}-\eqref{main2} have been introduced for future
convenience when $d=5$ and the equations rewritten in the spinor
form.}
\begin{eqnarray}
Dv_{a} &=& -2\Lambda\Phi_{ab}\e^b\,,\label{ads1}\\
D\Phi_{ab} &=& \ff12(v_a\e_b-v_b\e_a)\,.\label{ads2}
\end{eqnarray}
For any finite $\Lambda$ equations \eqref{ads1}-\eqref{ads2} are
equivalent to \eqref{Kil}-\eqref{CYK} upon the notation change
$v_a=-2\Lambda\xi_a\,, \Phi_{ab}=\xi_{ab}$ which gets degenerate
if $\Lambda=0$. Still, as was demonstrated, the system
\eqref{ads1}-\eqref{ads2} admits well defined flat limit. These
equations will be a starting point on the way of coordinate free
formulation of $5d$ black holes.

So far we have been considering generic $d$-dimensional case. Let
us restrict ourselves further on $d=5$ space-time, {\it i.e.}
vector indices range $a,b=0,\dots,4$. Most of the properties that
we need in what follows are easily derived from the spinor form of
\eqref{ads1}-\eqref{ads2}.

\section{Spinor analysis}\label{Spinor}
The vector $v_a$ has its antisymmetric traceless bispinor
counterpart $v_{\al\gb}=-v_{\gb\al}$ in the $5d$ spinorial
notation, while the antisymmetric tensor $\Phi_{ab}$ is
represented by the symmetric bispinor
$\Phi_{\al\gb}=\Phi_{\gb\al}$ (see appendix \ref{App-B}).
Analogously, f\"{u}nfbein $\e^a$ is given by the antisymmetric and
traceless bispinor 1-form $\e_{\al\gb}=-\e_{\gb\al}$. As a result,
equations \eqref{ads1}-\eqref{ads2} read
\begin{eqnarray}
Dv_{\al\gb}&=&-\ff{\Lambda}{2}(\Phi_{\al}{}^{\gga}\e_{\gga\gb}-\Phi_{\gb}{}^{\gga}\e_{\gga\al})\,,\label{main1}\\
D\Phi_{\al\gb}&=&\ff12(v_{\al}{}^{\gga}\e_{\gga\gb}+v_{\gb}{}^{\gga}\e_{\gga\al})\,.\label{main2}
\end{eqnarray}
This system has two independent Lorentz scalars which can choose
to be
\be\label{scal}
H=\sqrt{\det{\Phi_{\al\gb}}}\,,\quad
Q=\ff14\Phi_{\al\gb}\Phi^{\al\gb}\,.
\ee
For these, one finds
\be\label{dH}
dH=-\ff12H(\Phi^{-1})^{\al\gb}v_{\al}{}^{\gga}\e_{\gga\gb}\,,\quad
dQ= \ff12\Phi^{\al\gb}v_{\al}{}^{\gga}\e_{\gga\gb}\,.
\ee
The rest scalars are either expressed via these two through Fierz
identities \eqref{D5}, \eqref{D9}-\eqref{D11} (see Appendix
\ref{App-D}) or get reduced to the following first integrals
\begin{eqnarray}
P^2 &=& \ff14v_{\al\gb}v^{\al\gb}+\Lambda Q=v^2+\Lambda
Q=const\,,\label{I1}\\  I_1 &=&
-\ff12\Big(\ff14\Phi_{\al\gb}\Phi_{\gga\gd}v^{\al\gga}v^{\gb\gd}+
P^2\,Q-\ff{\Lambda}{2}(Q^2+H^2)\Big)=const\,,\label{I2}\\
I_2 &=& \ff{i}{4}(\Phi^2)_{\al\gb}v^{\al\gb}=const\,,\label{I3}
\end{eqnarray}
It is straightforward to check using \eqref{main1}-\eqref{main2}
that $dP^2=dI_{1,2}=0$. The number of first integrals is related
to the rank of space-time algebra which is equal to three for
either Poincar\'{e}, $so(4,2)$ or $so(5,1)$. One can think of the
fields $v_{\al\gb}$ and $\Phi_{\al\gb}$ as of the spinor
representation of space-time algebra satisfying the zero-curvature
condition \eqref{main1}-\eqref{main2}. The first integrals
\eqref{I1}-\eqref{I3} are hence the corresponding Casimir
operators. Their vector form amounts to
\begin{eqnarray}
P^2 &=& v_a v^a-\ff{\Lambda}{2} \Phi_{ab}\Phi^{ab}\,,\\
I_1 &=& 2(\Phi^2)_{ab}v^av^b+\ff14\Phi_{ab}\Phi^{ab}\cdot
P^2+\ff{\Lambda}{4}(Q^2+H^2)\,,\\
I_2 &=& \Phi_{ab}\Phi_{cd}v_e\varepsilon^{abcde}\,.
\end{eqnarray}
The system \eqref{main1}-\eqref{main2} possesses a number of
remarkable properties. One of the most important is its relation
to the solutions of massless field equations on $(A)dS_5$. The
fields $\Phi_{\al\gb}$ and $v_{\al\gb}$ can be shown to generate
the whole tower of integer spin solutions of $(A)dS_5$ Fronsdal
equations. Spin zero and spin one cases are easily derivable.
Indeed, from \eqref{dH} it is straightforward to obtain
\be\label{s=0}
\Box \ff{1}{H}=4\ff{\Lambda}{H}\,.
\ee
Hence, the field $\phi=\ff1H$ satisfy the Klein-Gordon equation.
The mass-like term on the r.h.s. of \eqref{s=0} exactly
corresponds to that of the massless scalar on $(A)dS_5$. Massless
spin one field takes its origin from a source free Maxwell tensor,
which can be constructed as follows. Consider
\be\label{Maxwell}
F_{\al\gb}=\ff1H\Phi^{-1}_{\al\gb}\,.
\ee
Using \eqref{main1}-\eqref{main2}, one finds
\be\label{maxwell}
dF_{\al\gb}=\ff{1}{2H}(F_{\al\gb}F_{\gga\gd}+
F_{\al\gga}F_{\gb\gd}+F_{\al\gd}F_{\gb\gga})v^{\gga\gl}\e_{\gl}{}^{\gd}\,.
\ee
The tensor $S_{\al\gb\gga\gd}=F_{\al\gb}F_{\gga\gd}+
F_{\al\gga}F_{\gb\gd}+F_{\al\gd}F_{\gga\gb}$ which appears in the
parenthesis of \eqref{maxwell} being totally symmetric entails the
identities
\be
D_{[\al\gb}F_{\gga]\gd}=0\,,\quad
D_{\al\gga}F_{\gb}{}^{\gga}-D_{\gb\gga}F_{\al}{}^{\gga}=0
\ee
both equivalent in vector notation to Maxwell equations for the
tensor $F^{ab}=\ff18\Gamma^{ab}_{\al\gb}F^{\al\gb}$:
\be\label{Maxeq}
\partial_{[a}F_{bc]}=0\,,\quad D_{b}F^{b}{}_{a}=0\,.
\ee
Therefore, locally
\be
F_{ab}=\partial_{a}A_{b}-\partial_{b}A_{a}\,,\quad \Box
A_a-D_bD_aA^b=0\,.
\ee
To get an appropriate spin one potential $A^a$ and proceed to
higher spins we note that these massless fields are uniformly
described by Kerr-Schild vectors.

\subsection{Kerr-Schild vectors and massless fields}\label{KS-massless}

In this section we generalize the construction of \cite{DMV} to
the five-dimensional case. The idea of \cite{DMV} was to generate
Kerr-Schild vectors out of the Killing vector $v^a$ by projecting
it onto appropriate light-like directions. These directions were
associated with eigen spinors of the projectors made of the
$\Phi_{ab}$ field. Ranks of so-defined projectors were equal to
one. The whole scheme was essentially four dimensional based on
two-component spinors leaving but little possibility for
higher-dimensional generalization. The desired generalization does
exist after all as we are going to demonstrate. Its key element
turns out to be Clifford algebra rather than two-component
spinors.

Consider the following projectors to single out light-like vectors
\be\label{proj}
\Pi^{\pm}_{\al\gb}=\ff12(\gep_{\al\gb}\pm X_{\al\gb})\,,
\ee
where $\gep_{\al\gb}=-\gep_{\gb\al}$ is the charge conjugation
matrix (see appendix \ref{App-B}), $X_{\al\gb}=X_{\gb\al}$ and
\be\label{idem}
(X^2)_{\al}{}^{\gb}=\gd_{\al}{}^{\gb}\,.
\ee
Their straightforward properties are
\be\label{pr-prop}
\Pi^{\pm}_{\al}{}^{\gga}\Pi^{\pm}_{\gga\gb}=\Pi^{\pm}_{\al\gb}\,,\quad
\Pi^{\pm}_{\al}{}^{\gga}\Pi^{\mp}_{\gga\gb}=0\,,\quad
\Pi^{+}_{\al\gb}=-\Pi^{-}_{\gb\al}\,.
\ee
From \eqref{pr-prop} it is obvious that the vectors
\[
v^{+}_{\al\gb}=\Pi^{+}_{\al\gga}\Pi^{+}_{\gb\gd}v^{\gga\gd}\,,\quad
v^{-}_{\al\gb}=\Pi^{-}_{\al\gga}\Pi^{-}_{\gb\gd}v^{\gga\gd}
\]
are both light-like. The involutory matrix \eqref{idem} is fixed
exactly once being dependent on $\Phi_{\al\gb}$ only. In that
case, one readily finds
\be\label{X}
X_{\al\gb}=\ff{1}{2r}(\Phi_{\al\gb}+H\Phi^{-1}_{\al\gb})\,,
\ee
where
\be\label{r}
r^2=\ff12(H-Q)\,.
\ee
The two Kerr-Schild vectors are given by
\be\label{KS}
k_{\al\gb}=\ff{v^{+}_{\al\gb}}{v^+v^-}\,,\qquad
n_{\al\gb}=\ff{v^{-}_{\al\gb}}{v^+v^-}\,,
\ee
where
\be
v^+v^-=\ff14v^{+}_{\al\gb}v^{-\al\gb}=\ff14\Pi^{+}_{\al\gga}\Pi^{+}_{\gb\gd}v^{\gga\gd}v^{\al\gb}\,.
\ee
The normalization is chosen such that $k^{a}v_{a}=n^{a}v_{a}=1$.
The vectors in \eqref{KS} are null by definition. That these
satisfy geodesity condition
\be\label{geodetic}
k^{a}D_{a}k_b=n^{a}D_{a}n_b=0
\ee
is far from being obvious and deserves some more attention.
Equation \eqref{geodetic} can be proven using
\eqref{main1}-\eqref{main2} and definitions \eqref{proj},
\eqref{X}, \eqref{r} by direct differentiation along with the use
of $5d$ Fierz identities. This calculation gives little perception
as to what extent the proposed construction is general, though. In
fact, property \eqref{geodetic} is valid for null vectors based on
similar projectors defined within generic Clifford algebra and
does not require an explicit form of the involutory matrix
$X_{\al\gb}$ for its proof. Aiming possible generalization of our
construction to higher dimensions, we provide the proof of
\eqref{geodetic} which does not refer to formula \eqref{X} in
appendix \ref{App-C}.

Equations \eqref{main1}-\eqref{main2} are invariant under the
discrete symmetry
\be\label{disc}
\tau_{c}: v^a\to c\cdot v^a\,,\quad \Phi_{ab}\to c\cdot
\Phi_{ab}\,,
\ee
where $c$ is an arbitrary real constant. It interchanges
Kerr-Schild vectors for $c=-1$
\be
k^a=\tau_{\!-1}\,(n^a)\,.
\ee
Properly normalized difference between the two Kerr-Schild vectors
is a total derivative. Indeed, it can be easily shown that
\be
d\big(\ff1H(k_{\al\gb}-n_{\al\gb})\e^{\al\gb}\Big)=0\,.
\ee
Moreover, both vectors appear as potentials for the Maxwell tensor
defined in \eqref{Maxwell}
\be\label{Maxpot}
4F_{\al\gb}=\partial_{\al\gga}\Big(\ff1Hk^{\gga}{}_{\gb}\Big)+
\partial_{\gb\gga}\Big(\ff1Hk^{\gga}{}_{\al}\Big)= \partial_{\al\gga}\Big(\ff1Hn^{\gga}{}_{\gb}\Big)+
\partial_{\gb\gga}\Big(\ff1Hn^{\gga}{}_{\al}\Big)\,,
\ee
or, equivalently, in the vector form
\be
4F_{ab}=\partial_{b}\ff{k_{a}}{H}-\partial_a\ff{k_{b}}{H}=
\partial_{b}\ff{n_{a}}{H}-\partial_a\ff{n_{b}}{H}\,.
\ee
From Maxwell equations \eqref{Maxeq} we, therefore, obtain
\be\label{s=1}
\Box\Big(\ff{k_a}{H}\Big)-D_bD_a\Big(\ff{k^b}{H}\Big)=0\,.
\ee
Equations \eqref{s=0} and \eqref{s=1} correspond to massless
scalar and spin $s=1$ equations on $(A)dS_5$, respectively. A
natural candidate for the massless spin $s=2$ field is
$h_{ab}=\ff1Hk_ak_b$. It is straightforward if somewhat involved
calculation based on \eqref{D6}-\eqref{D11} that confirms this
guess yielding linearized Einstein equations
\be\label{s=2}
\Box h_{mn}-D_{p}D_{m}h^{p}{}_{n}-D_{p}D_{n}h^{p}{}_{m}=-8\Lambda
h_{mn}\,.
\ee
A sequence of Kerr-Schild massless fields naturally goes on as is
clear from \eqref{s=0}, \eqref{s=1}, \eqref{s=2} and generates the
spin--$s$ Fronsdal field
\be
\phi_{a_1\dots a_s}=\ff1Hk_{a_1}\dots k_{a_s}
\ee
satisfying the $(A)dS_5$ Fronsdal equations
\be\label{s}
\Box\phi_{a_1\dots a_s}-sD_bD_{(a_1}\phi_{a_2\dots a_s)}{}^{b}=
-2\Lambda(s-1)(s+2)\phi_{a_1\dots a_s}
\ee
in agreement\footnote{To obtain literal agreement with the result
of Metsaev \cite{Metsaev} for the case of the equations for
totally symmetric Fronsdal fields in $AdS$, one needs to apply the
Ricci identity $[D_a,
D_b]T^c=\Lambda(\gd_{a}{}^{c}T_b-\gd_{b}{}^{c}T_{a})$ to change
the order of derivatives in the second term on the l.h.s of
\eqref{s}.} for $d=5$ with \cite{Metsaev}. To prove \eqref{s} it
is sufficient to prove it for $s=0,1,2$. The generic $s$ case
comes out as a straightforward consequence.

A comment is now in order. That geodesic shear-free congruence
generates solutions of massless equations is natural in four
dimensions as a consequence of the famous Penrose twistor
transform \cite{Penrose}. Indeed, the poles of the integrand of
the latter exactly reproduce Kerr-Schild congruences. To the best
of our knowledge no such explanation is known for higher
dimensions.

\section{Black holes}\label{BH}
In accordance with the considerations of Section \ref{K-S}, the
constructed null vectors \eqref{KS} have all necessary properties
to solve $d=5$ Einstein equations $R_{mn}=4\Lambda g_{mn}$ in the
form \eqref{KSvec}.  Any vector of \eqref{KS} can be chosen as a
Kerr-Schild vector in \eqref{KSvec}. The resulted metrics are
equivalent as will be demonstrated. Let us choose $k_{\al\gb}$ for
definiteness. So
\be\label{KSset}
g_{mn}=\bar{g}_{mn}+\ff{2M}{H}k_mk_n\,,
\ee
where $\bar{g}_{mn}$ is the $(A)dS_5$ metric, $M$ is an arbitrary
constant, $H$ is given by \eqref{scal} and $k_m$ is defined in
\eqref{KS}, solves Einstein equations with the cosmological
constant in coordinate independent fashion. Weyl tensor is
calculated to be
\be\label{Weyl1}
C_{\al\gb\gga\gd}=-\ff{32M}{H}(F_{\al\gb}F_{\gga\gd}+F_{\al\gga}F_{\gb\gd}+F_{\al\gd}F_{\gb\gga})\,,
\ee
its type is $\underline{22}$ according to the De Smet
classification \cite{DeSmet}.

Casimir invariants \eqref{I1}-\eqref{I3} are diffeomorphism
invariant characterizing a set of {\it inequivalent} metrics for
different values of constants $P^2, I_1, I_2$. Using the scale
ambiguity \eqref{disc} it is possible, however, to set {\it e.g.,}
$P^2$ discrete equal to either $P^2=-1,0,1$. An arbitrary constant
$M$ thus restores back scale ambiguity of fixed $P^2$, unless
$P^2=0$ in which case $M$ is no longer relevant and can be taken
$M=1$, for example. Eventually, the metrics \eqref{KSset} are
fully characterized by one discrete parameter $P^2$ and three
continuous $M, I_1, I_2$. Let us demonstrate now that
\eqref{KSset} is a five-dimensional analogue of $d=4$
Carter-Plebanski metric \cite{Carter, Pleb} without
electro-magnetic charges (note, that there are no NUT-charges in
five dimensions). It contains Myers-Perry black hole rotating
about two independent planes for $P^2=-1$. The black hole mass is
given by $M$ while its angular momenta are encoded in $I_1, I_2$.
The cases with $P^2=1$ and $P^2=0$ are novel in $d=5$ and can be
interpreted as ``tachyonic'' and ``light-like'' black holes,
respectively.

\subsection{Explicit realization}

To simplify the calculations, let us set $\Lambda=0$ to focus on
black holes in Minkowski space-time. Let us now enlist all of the
metrics explicitly in the Cartesian reference frame
$x^{a}=(t,x,y,z,u)$. To do so, we write down the general solution
of the equations \eqref{main1}-\eqref{main2}
\be
v_{\al\gb}=v^{0}_{\al\gb}=const\,,\quad
\Phi_{\al\gb}=\ff12(v_{\al}{}^{\gga}x_{\gga\gb}+v_{\gb}{}^{\gga}x_{\gga\al})+\Phi^0_{\al\gb}\,,\quad
\Phi^{0}_{\al\gb}=const\,,
\ee
where $x_{\al\gb}=x^a\gga_{a\,\al\gb}$. Using a convenient
parametrization, all inequivalent solutions can be summarized in
the following table
\begin{center}
\begin{tabular}{|l|l|l|c|c|c|} \hline
Type     &  Killing vector $v_{\al\gb}$ & Lorentz generator
$\Phi^{0}_{\al\gb}$ & $P^2$ & $I_1$ & $I_2$
\\ \hline {Kerr}    & $\frac{\partial}{\partial t}$ & $a\Gamma^{xy}_{\al\gb}+b\Gamma^{zu}_{\al\gb}$
& $-1$ & $b^2+a^2$ & $2ab$
 \\
{Light-like Kerr }    & $\frac{\partial}{\partial
t}+\frac{\partial}{\partial x}$ &
$a\Gamma^{xy}_{\al\gb}+b\Gamma^{zu}_{\al\gb}$ &
        $0$ & $a^2$ & $2ab$ \\
{Tachyonic Kerr}    & $\frac{\partial}{\partial x}$ &
$a\Gamma^{ty}_{\al\gb}+b\Gamma^{zu}_{\al\gb}$ &  $+1$
        & $a^2-b^2$ & $2ab$ \\ \hline
\end{tabular}
\end{center}

\centerline{{\bf Table 1.} Classification of Kerr-Schild solutions
on 5d Minkowski space by its Poincar\'{e} invariants}

\subsection*{Kerr metric, $P^2=-1$}
To reproduce a Kerr black hole we choose $v^{a}=(1,0,0,0,0)$ or
equivalently $v_{\al\gb}=\gga_{0\al\gb}$ and
\be
\Phi^0_{\al\gb}=a\Gamma^{xy}_{\al\gb}+ b\Gamma^{zu}_{\al\gb}\,,
\ee
where $x^{a}=(t,x,y,z,u)$. A straightforward calculation gives
\be
\det\Phi_{\al\gb}=H^2\,,
\ee
where
\be
H=\ff{1}{r^2}(r^2+a^2)(r^2+b^2)\Big(1-\ff{a^2(x^2+y^2)}{(r^2+a^2)^2}-\ff{b^2(z^2+u^2)}{(r^2+b^2)^2}\Big)
\ee
and $r$ defined in \eqref{r} reads
\be
\ff{x^2+y^2}{r^2+a^2}+\ff{u^2+z^2}{r^2+b^2}=1\,.
\ee
Casimir invariants are
\be
P^2=-1\,,\quad I_1=a^2+b^2\,,\quad I_2=2ab\,.
\ee
Thus,
\be
a=\ff12\Big(\sqrt{I_1+I_2}+\sqrt{I_1-I_2}\Big)\,,\quad
b=\ff12\Big(\sqrt{I_1+I_2}-\sqrt{I_1-I_2}\Big)\,,
\ee
For the Kerr-Schild vector $k_a$ we also find
\begin{eqnarray}\label{kvec}
k_0=1\,,\quad k_1=-\ff{xr+ay}{r^2+a^2}\,,\quad
k_2=-\ff{yr-ax}{r^2+a^2}\,,\\
k_3=-\ff{zr+bu}{r^2+b^2}\,,\quad k_4=-\ff{ur-bz}{r^2+b^2}\,.
\end{eqnarray}
This case corresponds to the familiar $5d$ Myers-Perry metric
\cite{MP}. The parameters $a$ and $b$ are the angular momenta per
unit mass of a rotating black hole about the $xy$-plane and the
$zu$-plane. Coordinates of the another Kerr-Schild vector $n_a$
are reproduced upon sign flip $(a,b,x,y,z,u\to-a,-b,-x,-y,-z,-u)$
in \eqref{kvec}.

\subsection*{Light-like Kerr, $P^2=0$}
Another solution with $P^2=0$ which can be called ``light-like''
Kerr has its $4d$ analog described by the Carter-Plebanski metric
\cite{Carter, Pleb} with the discrete Carter parameter $\gep=0$.
The value $P^2=0$ can be reached by the vector field
\be
v^a=(1,1,0,0,0)\,.
\ee
Taking
\be
\Phi^0_{\al\gb}=a\Gamma^{xy}_{\al\gb}+ b\Gamma^{zu}_{\al\gb}\,,
\ee
we obtain
\be
P^2=0\,,\quad I_1=a^2\,,\quad I_2=2ab\,.
\ee
Let us note that unlike Kerr, in this case $I_1=0$ implies
$I_2=0$. The equations \eqref{r} now reads
\be
a^2\ff{u^2+z^2}{r^2+b^2}+(t-x)^2-(y+a)^2=r^2-y^2\,.
\ee
The function $H$ and the Kerr-Schild vector $k_a$ are
\be
H=(r^2+b^2)\Big(1+a^2\ff{u^2+z^2}{(r^2+b^2)^2}\Big)\,,
\ee
\begin{eqnarray}
k_0=1+\ff{r^2+ay-r(t-x)}{a^2}\,,\quad
k_1=\ff{-r^2-ay+r(t-x)}{a^2}\,,\\
k_2=\ff{r+x-t}{a}\,,\quad k_3=-\ff{bu+zr}{r^2+b^2}\,,\quad
k_4=\ff{bz-ur}{b^2+r^2}\,.
\end{eqnarray}
The physical interpretation of this solution is not
straightforward. Its four dimensional analog suffers from
pathalogies in global properties \cite{GrifPodol}, so we do not
expect that the $5d$ counterpart would be any better. The notion
``light-like'' is justified, however, by the asymptotic behavior
of the metric. Unlike the Kerr case where the gravitational field
at large distances is described by a point source on a time-like
surface, the present case asymptotically corresponds to a
light-like source.

\subsection*{Tachyonic Kerr, $P^2=1$}
This case is naturally described by
\be
v^a=(0,1,0,0,0)\,,\quad
\Phi^{0}_{\al\gb}=a\Gamma^{ty}_{\al\gb}+b\Gamma^{zu}_{\al\gb}\,.
\ee
The corresponding Poincar\'{e} invariants are
\be
P^2=1\,,\quad I_1=a^2-b^2\,,\quad I_2=2ab\,.
\ee
The scalars $r$, $H$ and Kerr-Schild vector $k_a$ take the
following form
\be
\ff{u^2+z^2}{r^2+b^2}+\ff{y^2-t^2}{r^2-a^2}=-1\,,
\ee
\be
H=\ff{1}{r^2}(r^2-a^2)(r^2+b^2)\Big(1-\ff{a^2(y^2-t^2)}{(r^2-a^2)^2}
+\ff{b^2(u^2+z^2)}{(r^2+b^2)^2}\Big)\,,
\ee
\begin{eqnarray}
k_0=\ff{-ay+rt}{r^2-a^2}\,,\quad k_1=1\,,\quad
k_2=\ff{at-yr}{r^2-a^2}\,,\\
k_3=-\ff{bu+zr}{r^2+b^2}\,,\quad k_4=\ff{bz-ur}{r^2+b^2}\,.
\end{eqnarray}
This solution has a $4d$ analog represented by the non-charged
Carter-Plebanski metric with $\gep=-1$. Its physical relevance is
not clear and it has some pathologies in the global structure
\cite{GrifPodol}. Asymptotically metric \eqref{KSset} describes
the gravitational field of a tachyonic source.

As it was already mentioned, the parameter $P^2$ is analogous to
$d=4$ Carter-Plebanski parameter $\gep$. For $\Lambda\neq 0$ the
proper normalization for Casimir invariants will be given in the
next section.

\subsection{Unfolded form of the black hole}\label{BHUS}

In section \ref{BH}, we have shown how black holes are generated
by a single $(A)dS_5$ global symmetry parameter in arbitrary
coordinates. Apart from its coordinate independence the advantage
of the proposed approach is that the fields which constitute black
hole's metric satisfy the background $(A)dS_5$ equation. The
latter is just the zero curvature condition of $so(4,2)$ or
$so(5,1)$ algebra thus being pure gauge. Black holes appear as the
Kerr-Schild transform of the background metric \eqref{KSset}. It
is possible, however, to deform \eqref{main1}-\eqref{main2} such
that its consistency condition $\D^2\sim R$, $\D R=0$ would imply
black hole's curvature rather than $(A)dS_5$. Indeed, consider the
following one-parametric deformation of
\eqref{main1}-\eqref{main2}
\begin{eqnarray}
\D
v_{\al\gb}&=&-\ff{\Lambda}{2}(\Phi_{\al}{}^{\gga}\e_{\gga\gb}-\Phi_{\gb}{}^{\gga}\e_{\gga\al})+
\ff{\mu}{H}(\Phi^{-1}_{\al}{}^{\gga}\e_{\gga\gb}-\Phi^{-1}_{\gb}{}^{\gga}\e_{\gga\al})\,,\label{bhus1}\\
\D\Phi_{\al\gb}&=&\ff12(v_{\al}{}^{\gga}\e_{\gga\gb}+v_{\gb}{}^{\gga}\e_{\gga\al})\,,\label{bhus2}
\end{eqnarray}
where $\mu$ is an arbitrary real constant and
$H=\sqrt{\det{\Phi_{\al\gb}}}$. This system is consistent,
provided the Weyl tensor in \eqref{weyl} is given by
\be
C_{\al\gb\gga\gd}=-\ff{32\mu}{H^3}(\Phi^{-1}_{\al\gb}\Phi^{-1}_{\gga\gd}+\Phi^{-1}_{\al\gga}
\Phi^{-1}_{\gb\gd}+\Phi^{-1}_{\al\gd}\Phi^{-1}_{\gb\gga})\,.
\ee
It is straightforward to check that the Bianchi identity $\D
C_{\al\gb\gga\gd}\wedge\e^{\gga\gl}\wedge\e_{\gl}{}^{\gd}=0$
holds. Equations \eqref{bhus1}-\eqref{bhus2} have clearly the
unfolded form\footnote{Recall, the dynamical system is said to
have the unfolded form if it is formulated in terms of first order
differential equations for differential $p$-forms $W^{A}$, $p\geq
0$ and generalized curvatures $R^A=dW^A+F^A(W)$, provided the
functions $F^A$ are subject to the generalized Jacobi identities
$F^B\wedge\ff{\gd F^A}{\gd W^B}=0$ that guarantee the generalized
Bianchi identities $dR^A=R^B\wedge\ff{\gd F^A}{\gd W^B}$ (see {\it
e.g.,} \cite{actions}).} as being consistent within given set of
fields. Note, the equation \eqref{bhus2} remains unchanged as
compared to \eqref{main2}.

Analogously to \cite{DMV}, the deformed system
\eqref{bhus1}-\eqref{bhus2} shares many properties with the
nondeformed one \eqref{main1}-\eqref{main2}. Particularly, as
follows from \eqref{bhus1}, $v_{\al\gb}$ is still a Killing
vector. The field $\Phi_{\al\gb}$ remains a principal CYK as well.
It has the same number of first integrals. These are given by
\begin{eqnarray}
I_0 &=& v^2+\Lambda Q-\ff{2\mu}{H}\,,\label{c1}\\
I_1 &=&
-\ff12\Big(\ff14\Phi_{\al\gb}\Phi_{\gga\gd}v^{\al\gga}v^{\gb\gd}
+I_0\,Q-\ff{\Lambda}{2}(Q^2+H^2)\Big)+\mu\,,\label{c2}\\
I_2 &=& \ff{i}{4}(\Phi^2)_{\al\gb}v^{\al\gb}\label{c3}\,,
\end{eqnarray}
Note that the dimension of the mass parameter $\mu$ in $d=5$ is
$[\mu]=2$ and so is the dimension of $I_{1,2}$. Hence, $\mu$ can
be added up to arbitrary factors to $I_1$ and $I_2$. A choice of
that particular factor in $I_1$ will become clear later on.

A space-time described by the equations
\eqref{bhus1}-\eqref{bhus2} can be identified as follows.
According to \cite{Frolov}, Einstein spaces that admit the
principal CYK tensor \eqref{bhus2} should be of Chen-L\"{u}-Pope
type black holes \cite{CLP}. In our five-dimensional case these
were classified in the Table 1 for $\Lambda=0$. The type of
solution is defined by the sign of $I_0$ invariant -- Myers-Perry
for $I_0<0$, ``light-like'' for $I_0=0$ and ``tachyonic'' for
$I_0>0$. Indeed, by appropriate global rescaling $v\to c v$,
$\Phi\to c\Phi$, $\mu\to c^4\mu$ $I_0$ can always be set to either
$-1,0,1$. Hence, the sign of $I_0$ distinguishes between
inequivalent solutions.

Let us enlist some more properties of \eqref{bhus1}-\eqref{bhus2}.
It admits at least three Killing vectors which arise due to the
existence of the CYK field $\Phi_{\al\gb}$. The first one,
$v_{\al\gb}$, is manifest. To identify the other two, one
constructs the Killing tensor $K_{ab}=K_{ba}$ which produces the
latter as $\xi^{(1)}_{a}=K_{ab}v^b$ and
$\xi^{(2)}_{a}=K_{ab}\xi^{(1)\, b}$ (see \cite{Frolov} for more
details). In the spinor notation the Killing tensor is represented
by the following multispinor
\[
K_{\al\gb,\, \gga\gd}=K_{\gga\gd,\, \al\gb}\,,\quad K_{\al\gb,\,
\gga\gd}=-K_{\gb\al,\, \gga\gd}=-K_{\al\gb,\, \gd\gga}\,,\quad
K_{\al}{}^{\al}{}_{,\, \gga\gd}=K_{\al\gb,\,\gga}{}^{\gga}=0\,.
\]
In terms of CYK field $\Phi_{\al\gb}$, it has the following
explicit form:
\be
K_{\al\gb,\,
\gga\gd}=\Phi_{\al\gga}\Phi_{\gb\gd}-\Phi_{\gb\gga}\Phi_{\al\gd}+\ff12(\Phi^2_{\gga\gd}\gep_{\al\gb}
+\Phi^2_{\al\gb}\gep_{\gga\gd})+\ff18\Phi_{\mu\nu}\Phi^{\mu\nu}\gep_{\al\gb}\gep_{\gga\gd}\,.
\ee
The corresponding Killing vectors then read
\be
\xi_{\al\gb}^{(1)}=2\Phi_{\al\gga}\Phi_{\gb\gd}v^{\gga\gd}+
\ff12\Phi^2_{\gga\gd}v^{\gga\gd}\gep_{\al\gb}\,,\quad
\xi_{\al\gb}^{(2)}=2\Phi_{\al\gga}\Phi_{\gb\gd}\xi^{(1)\,\gga\gd}+
\ff12\Phi^2_{\gga\gd}\xi^{(1)\,\gga\gd}\gep_{\al\gb}\,.
\ee

Recall now, that the $(A)dS_5$ equations
\eqref{main1}-\eqref{main2} admit the source-free Maxwell tensor
\eqref{Maxwell}. Introduction of the deformation parameter $\mu$
leaves it unaffected, {\it i.e.,} equations \eqref{Maxwell} and
\eqref{maxwell} still hold. Same is the story with Kerr-Schild
vectors. The whole projector construction and its properties
\eqref{proj}-\eqref{geodetic} extend to black hole unfolded system
upon the change $D\to\D$ in \eqref{geodetic}. Pretty much as in
the nondeformed case, Kerr-Schild vectors defined in \eqref{KS}
generate Maxwell potentials \eqref{Maxpot} resulting in the
following Maxwell equations
\be\label{BHmax}
\Box\Big(\ff{k_a}{H}\Big)-\D_b\D_a\Big(\ff{k^b}{H}\Big)=0\,.
\ee
One can equally well replace $k^a$ with $n^a$ in \eqref{BHmax}.
Interestingly enough, even linearized Einstein equations
\eqref{s=2} hold upon covariantization $D\to\D$. This can not be
said about Fronsdal equations \eqref{s} for $s>2$ which naturally
break down for $\mu\neq 0$. Technically, \eqref{s} would have been
held had equation \eqref{s=0} been satisfied. This is not the
case; instead for $\mu\neq 0$ one finds
\be
\Box\ff1H=4\ff{\Lambda}{H}+8\mu\ff{Q}{H^4}\,,
\ee
where $Q$ is defined in \eqref{scal}.

From algebraic standpoint, both the $(A)dS_5$ and black hole
systems are to much extent equivalent and coincide at $\mu=0$. The
equivalence becomes manifest upon field redefinition that maps one
system to another. An instructive way to obtain this map is to
perform an integrating flow $\ff{\partial}{\partial\mu}$ from
initial surface $\mu=0$ to some fixed finite value. The flow in
general is governed by first order ordinary differential equations
of the form
\be\label{flow}
\ff{\partial}{\partial\mu}
v_{\al\gb}=f^{(1)}_{\al\gb}(v,\Phi,\mu)\,,\quad
\ff{\partial}{\partial\mu}\Phi_{\al\gb}=f^{(2)}_{\al\gb}(v,\Phi,\mu)\,,\quad
\ff{\partial}{\partial\mu}\e_{\al\gb}=f^{(3)}_{\al\gb}(\e,v,\Phi,\mu)\,,
\ee
where the functions $f^{(i)}$ to be fixed by the integrability
requirement
\be\label{flow-cons}
[d, \ff{\partial}{\partial\mu}]=0\,,
\ee
where $d$ -- is the space-time differential. Solving the evolution
equations \eqref{flow} one finds the searched for map. This
strategy has been accomplished for $d=4$ black holes in
\cite{DMV}. Using similar approach we propose the following
integrating flow for $d=5$ black holes
\be\label{fl}
\ff{\partial}{\partial\mu}\Phi_{\al\gb}=0\,,\quad
\ff{\partial}{\partial\mu} v_{\al\gb}=\ff1H k_{\al\gb}\,,\quad
\ff{\partial}{\partial\mu}\e_{\al\gb}=\ff{1}{4H}k_{\al\gb}k_{\gga\gd}\e^{\gga\gd}\,.
\ee
Equations \eqref{fl} are motivated by the Kerr-Schild shift and
turn out to be consistent with \eqref{flow-cons}. They can be
easily integrated if one notices that the Kerr-Schild vector $k^a$
is constant along the flow. Indeed, applying \eqref{fl} to its
definition \eqref{KS} we obtain
$\ff{\partial}{\partial\mu}k_{\al\gb}=0$. Note, that this is not
true for $n^a$, {\it i.e.,} $\ff{\partial}{\partial\mu}
n_{\al\gb}\neq 0$. As a result, the solution of \eqref{fl} reads
\be
\Phi_{\al\gb}=\Phi_{\al\gb}^0\,,\quad
v_{\al\gb}=v_{\al\gb}^0+\ff{\mu}{H}k_{\al\gb}^0\,,\quad
\e_{\al\gb}=\e_{\al\gb}^0+\ff{\mu}{4H}k_{\al\gb}^0 k_{\gga\gd}^0
\e^{0\,\gga\gd}\,,
\ee
where the subscript $0$ is assigned to $\mu=0$ nondeformed fields
of \eqref{main1}-\eqref{main2}. Recall that $k_a=k^0_a$. The first
integrals \eqref{c1}-\eqref{c3} are invariant along the flow as
well $\ff{\partial}{\partial\mu}I_{0,\,1,\,2}=0$, thanks to the
additional $\mu$ added to \eqref{c2}:
\be
I_{0,\,1,\,2}=I_{0,\,1,\,2}^0\,.
\ee
Black hole parameters such as angular momenta and the
Carter-Pleba\'{n}ski parameter are therefore encoded in Casimir
invariants \eqref{I1}-\eqref{I3}. The metric can be now easily
calculated
\be
ds^2=\ff14\e_{\al\gb}\cdot\e^{\al\gb}=ds_0^2+\ff{2\mu}{H}k_{m}k_{n}dx^m
dx^n\,.
\ee
For the physically important solution obtained by Hawking, Hunter
and Taylor-Robinson \cite{HHTR} that is for $d=5$ Myers-Perry
black hole in the presence of a non-zero cosmological constant,
the identification of the first integrals \eqref{c1}-\eqref{c3}
along with the mass parameter $\mu$ is found to be as follows
\begin{eqnarray}
I_0 &=& -1+\Lambda(a^2+b^2)\,,\label{hawk}\\
I_1 &=& a^2+b^2-\Lambda a^2b^2\,,\\
I_2 &=& 2ab\,,\\
\mu &=& M\,.\label{hawk2}
\end{eqnarray}
In deriving this result, we have compared the horizon equation
$\Delta(r)=0$ of \cite{HHTR} to the horizon condition in our
approach. To find the latter, we note that the scalar product of
two geodesic light-like congruences is degenerate on the horizon.
Hence, it is sufficient to analyze scalar product of two
Kerr-Schild vectors \eqref{KS} defined for
\eqref{bhus1}-\eqref{bhus2} which is given by
\be
k_a n^a=\ff{1}{v^+v^-}=\ff{2H}{\Delta_r}\,,
\ee
where
\be
\Delta_r=I_0r^2+2\mu+\Lambda r^4-I_1-\ff{I_2^2}{4r^2}\,.
\ee
Comparing $\Delta_r$ with that of \cite{HHTR}, one reproduces
\eqref{hawk}-\eqref{hawk2}.

\section{Conclusion}\label{conc}
Five-dimensional Myers-Perry type black hole have been
reconsidered in a coordinate independent way based on unfolded
description of dynamical systems. In addition to the cosmological
constant that was accounted for by Hawking, Hunter, and
Taylor-Robinson \cite{HHTR}, we also include the analogue of the
$4d$ discrete Carter-Plebanski parameter to the solution.
So-defined black holes acquire a beautiful algebraic
classification. The whole class is generated by a background
space-time (either $(A)dS_5$ or Minkowski) single global symmetry
parameter. Three Casimir invariants $P^2, I_1, I_2$ associated
with that parameter distinguish between inequivalent black holes
and produce black hole's ``hair'' -- the mass and angular momenta.
The classification resembles that of relativistic fields with
spins and masses originated from Casimir operators of
$AdS$/Poincar\'{e} algebra.  Ordinary black holes naturally arise
this way for $P^2<0$ along with the light-like ($P^2=0$) and
tachyonic ($P^2>0$). The value of $P^2$ can be associated with the
$5d$ analogue of the Carter-Plebanski parameter. The other two
invariants $I_1, I_2$ determine angular momenta. In this respect,
it should be noted that some indication that black holes can be
treated within representation theory was given in \cite{IS2008} at
the linearized level. That black holes are generated by a
background global symmetry parameter was shown in \cite{DMV} for
$d=4$ black holes. The analysis of \cite{DMV}, however, leaves no
indication if the proposed classification extends to higher
dimensions or not. Now when it is shown that it actually does for
$d=5$, we believe it could be extended to arbitrary dimension with
the use of Clifford algebra. Indeed, the proposed construction is
based on Kerr-Schild ansatz which mysteriously works for
Myers-Perry black holes, while the very Kerr-Schild vector arises
from spinor projectors as demonstrated.

Parallel to the result of \cite{DMV} where the unfolded form of
the $4d$ Carter-Plebanski black hole has been given as a
deformation of $AdS_4$ global symmetry condition, a similar
one-parametric deformation takes place for $d=5$ rotating black
holes without electro-magnetic charges. The deformed equations are
related to the vacuum ones via the integrating flow describing
evolution with respect to black hole mass. The integration of the
flow equations with the initial data corresponding to $(A)dS_5$
space allows us to express black hole fields that describe its
metric in terms of their vacuum values. It should be stressed that
in both four and five dimensions the crucial element that makes
deformation possible is the existence of a principal conformal
Killing-Yano tensor\footnote{In \cite{DMV} black hole deformation
was carried out in terms of Killing and Maxwell fields. The latter
is related to principal CYK.} for Myers-Perry type black holes.
Moreover, it is this field attributed to all higher-dimensional
Myers-Perry type black holes that makes variable separation for
Klein-Gordon and Dirac equations in black hole background possible
\cite{Frolov}.

Apart from elucidating the structure of $d=5$ black holes, one of
the goals of the present research was an application of higher
spin gauge theory machinery to describe black holes as it
hopefully allows us to generalize the latter to include
higher-spin interactions. In this respect, the analogous $4d$
result of \cite{DMV} turned out to be very instructive and indeed
allowed us to find a $4d$ higher-spin black hole solution
\cite{DV}. We hope it will be possible to generalize the $5d$
black hole in higher-spin theory as well. Indeed, the developed
approach based on a background global symmetry parameter provides
Kerr-Schild solutions for all free massless fields rather than
$s=2$ gravity only. The interaction between these massless fields
can in principal be accounted by the nonlinear higher-spin
equations \cite{HSd}. Little chance to do it straightforwardly
though, since the spinor form of these equations in $d=5$ is
lacking as yet.

The other motivation to focus on $d=5$ was the fact that it is the
minimal space-time dimension where black holes of nonspherical
topology arise \cite{bring}. Unfortunately the results of the
paper give no hint on whether black rings admit a similar
description based on a single global symmetry parameter of
Minkowski space-time. One thing is for certain, even if they
really do, their unfolded equations would be completely different
as compared to those considered in our work. From the algebraic
point of view, the black ring essentially differers from the
spherical black hole. Particularly, its Weyl tensor is not
algebraically special and it does not admit the principal CYK
field. Still one may ask oneself if there is a Ricci flat
consistent deformation of the Minkowski global symmetry condition
within the same set of fields. $5d$ spinor language seems well
adopted to tackle this problem.

\section*{Acknowledgement}
I'm indebted to K.B. Alkalaev, C. Iazeolla, P. Sundell and
especially to M.A. Vasiliev for useful discussions. This research
was supported in part by RFBR Grant No 11-02-00814 and Russian
President Grant No 5638.

\renewcommand{\theequation}{A.\arabic{equation}}
\renewcommand{\thesubsection}{A}
\renewcommand{\thesection}{Appendix}
\makeatletter \@addtoreset{equation}{subsection} \makeatother

\section{}

\subsection{Cartan formalism} \label{App-A}
We use the following definition for the Riemann tensor
\be
[D_m, D_n]T^k=R^{k}{}_{p,mn}T^p
\ee
or in terms of Christoffel symbols
\be
R^{k}{}_{l,mn}=\partial_m\Gamma^{k}_{ln}-\partial_{n}\Gamma^{k}_{lm}+
\Gamma^{p}_{ln}\Gamma^{k}_{pm}-\Gamma^{p}_{lm}\Gamma^{k}_{pn}
\ee
which are given by
\be
\Gamma^{l}_{mn}=\ff12g^{lp}(\partial_{n}g_{mp}+\partial_{m}g_{np}-\partial_{p}g_{mn})\,,
\quad D_nT^m=\partial_n T^m+\Gamma^{m}_{np}T^p\,.
\ee
The vacuum Einstein equations in five dimensions $(m,n=0,\dots,
4)$ read
\be\label{Ein}
R_{mn}=4\Lambda g_{mn}\,.
\ee
A convenient way to describe space-time geometry is to use the
Cartan formalism. Since it is used it in our analysis, we recall
it somewhat in detail. To proceed, introduce the antisymmetric
Lorentz connection one-form $dx^m\w_{ab,\,m}=-dx^m\w_{ba,\,m}$ and
f\"{u}nfbein one-form $dx^m\e_{ab,\,m}$, where fiber indices range
$a,b=0,\dots, 4$. Flat indices are raised and lowered by mostly
plus Minkowski metric $\eta_{ab}$. Cartan-Maurer equations have
the form
\begin{eqnarray}
\R_{ab} &=& d\w_{ab}+\w_{a}{}^{c}\wedge\w_{cb}\,,\label{Cartan1}\\
\R_a &=& \D\e_a=d\e_a+\w_{a}{}^{b}\wedge\e_b=0\,,\label{Cartan2}
\end{eqnarray}
where $\R_{ab}=\R_{ab,mn}dx^m\wedge dx^n$ is the curvature
two-form to be identified with the Riemann tensor as follows
\be
\R_{ab}=\ff12R_{mn,\,kl}\e^{m}_{a}\e^{n}_{b}dx^k\wedge dx^l\,,
\ee
provided the frame field $\e_a$ defines the metric
$g_{mn}=\e_{a,\,m}\e_{b,\,n}\eta^{ab}$. Equation \eqref{Cartan2}
is the metric postulate that sets torsion two-form $\R_a$ to zero.
The integrability condition $d^2=0$ for
\eqref{Cartan1}-\eqref{Cartan2} amounts to Bianchi identities
\be\label{int}
\D^2T_a=\R_{ab}T^b\,,\quad \D\R_{ab}\wedge\e^b=0\,,
\ee
where $T^a$ is an arbitrary vector. For $(A)dS_5$ space-time, for
example, the curvature two-form is $R_{ab}=\Lambda\e_a\wedge\e_b$.
Einstein equations \eqref{Ein} imply that curvature is equal to
that of $(A)dS_5$ up to a totally traceless tensor
\be\label{Weyl}
\R_{ab}=\Lambda\e_a\wedge\e_b+\ff12C_{ab,\,cd}\,\e^c\wedge\e^d\,,
\ee
where $C_{ab,\,cd}$ is the traceless Weyl tensor written in fiber
components. The integrability condition for Einstein spaces is,
therefore,
\be\label{int2}
\D^2T_a=\Lambda\e_a\wedge\e_bT^b+\ff12C_{ab,\,cd}T^b\e^c\wedge\e^d\,,\quad
\D C_{ab,\,cd}\,\e^b\wedge\e^c\wedge\e^d=0\,.
\ee

\renewcommand{\theequation}{B.\arabic{equation}}
\renewcommand{\thesubsection}{B}
\renewcommand{\thesection}{Appendix}
\makeatletter \@addtoreset{equation}{subsection} \makeatother
\subsection{Spinors in five dimensions} \label{App-B}

Consider Clifford algebra in five dimensions
\be
\gga_a\gga_b+\gga_b\gga_a=2\eta_{ab}\,.
\ee
Its dimension is $2^{[5/2]}=4$ and hence $5d$ $\gga$-matrices can
be realized by four dimensional as follows
\be
\gga_a=(\gga_{\hat{a}}, i\gga_5)\,,
\ee
where $\hat{a}=0,\dots, 3$ and $\gga_5=\gga_0\gga_1\gga_2\gga_3$.
A convenient choice for $4d$ $\gamma$-matrices is the Majorana
representation
\be
\gamma_0=\begin{pmatrix} -i \sigma^2 & 0 \\ 0 & i \sigma^2
\end{pmatrix}, \ \gamma_1=\begin{pmatrix} \sigma^1 & 0 \\ 0 &
\sigma^1 \end{pmatrix}, \ \gamma_2=\begin{pmatrix} \sigma^3 & 0 \\
0 & \sigma^3 \end{pmatrix}, \ \gamma_3=\begin{pmatrix} 0 & i
\sigma^2  \\ -i \sigma^2 & 0 \end{pmatrix},
\ee
\be
\gga_4=i\gga_0\gga_1\gga_2\gga_3=\begin{pmatrix} 0 & \gs^2 \\
\gs^2 & 0 \end{pmatrix}
\ee
where $\sigma^{1,2,3}$ are Pauli matrices. Restoring spinor
indices, we use the $\gga_{a\,\al}{}^{\gb}$-notation for
$\gamma$-matrices, $\al, \gb=1,\dots,4$. The charge conjugation
matrix $\gep_{\al\gb}=-\gep_{\gb\al}$ is antisymmetric and is used
to raise and lower spinor indices according to the rule
\be
\gga_{a\,\al\gb}=\gga_{a\,\al}{}^{\gd}\gep_{\gd\gb}\,,\quad
\gga_{a}^{\al\gb}=\gga_{a\,\gd}{}^{\gb}\gep^{\al\gd}\,,\quad
\gep_{\al\gd}\gep^{\gb\gd}=\gd_{\al}{}^{\gb}\,.
\ee
$\gga$-matrices are antisymmetric with respect to charge
conjugation, that is, $\gga_{a\,\al\gb}=-\gga_{a\,\gb\al}$.
Irreducibility, in addition, implies tracelessness condition
$\gga_{a,\,\al}{}^{\al}=0$. Thus, the number of spinorial
components of the traceless antisymmetric matrix
$\gga_{a\,\al\gb}$ is equal to $\ff{4(4-1)}{2}-1=5$ coinciding
with the number of components of the $5d$ vector. This makes it
possible to assign a vector to any antisymmetric and traceless
bispinor and vise versa
\be
T_{\al\gb}=T^a\gga_{a\,\al\gb}\,,\quad
T^a=\ff14\gga^{a}_{\al\gb}T^{\al\gb}\,.
\ee
Consider now spinor representation for Lorentz generators
\be
(\Gamma_{ab})_{\al\gb}=\ff12(\gga_a\gga_b-\gga_b\gga_a)_{\al\gb}\,.
\ee
These are symmetric $\Gamma^{ab}_{\al\gb}=\Gamma^{ab}_{\gb\al}$
and for fixed $a$ and $b$ have 10 spinor components as well as for
fixed $\al$ and $\gb$ there are 10 vector components. Using now
that the following antisymmetrized tensor product
$\gga^{a}_{\al\gb}\gga^{b}_{\gga\gd}-\gga^{b}_{\al\gb}\gga^{a}_{\gga\gd}$
has again the same number of components for either fixed vector or
spinor indices one easily arrives at the identity
\be\label{prop}
\gga^{a}_{\al\gb}\gga^{b}_{\gga\gd}-\gga^{b}_{\al\gb}\gga^{a}_{\gga\gd}=
\gep_{\al\gga}\Gamma^{ab}_{\gb\gd}-\gep_{\gb\gga}\Gamma^{ab}_{\al\gd}-
\gep_{\al\gd}\Gamma^{ab}_{\gb\gga}+\gep_{\gb\gd}\Gamma^{ab}_{\al\gga}\,,
\ee
which allows us to establish one to one correspondence between
antisymmetric tensor $T_{ab}=-T_{ba}$ and its symmetric bispinor
counterpart $T_{\al\gb}=T_{\gb\al}$:
\be
T_{\al\gb}=\Gamma^{ab}_{\al\gb}T_{ab}\,,\quad
T^{ab}=\ff18\Gamma^{ab}_{\al\gb}T^{\al\gb}\,.
\ee
The advantage of $5d$ spinors becomes especially vivid when the
Weyl tensor $C_{ab,\,cd}$ is concerned. Being a window-like
traceless diagram in tensorial terms, it has $35$ independent
components. Converting it with two $\Gamma$'s one obtains
\be\label{Ws}
C_{ab,cd}\Gamma^{ab}_{\al\gb}\Gamma^{cd}_{\gga\gd}=C_{(\al\gb\gga\gd)}+\dots\,,
\ee
where we have extracted a totally symmetric multispinor on the
r.h.s. of \eqref{Ws} such that $\dots$ contain the other
irreducible parts of the spinor decomposition. However, the number
of components of totally symmetric $C_{\al\gb\gga\gd}$ is equal to
$\ff{4\cdot 5\cdot 6\cdot 7}{4!}=35$. As a result, the $5d$ Weyl
tensor is completely represented by the totally symmetric
multispinor $C_{\al\gb\gga\gd}$
\be
C_{\al\gb\gga\gd}=C_{ab,cd}\Gamma^{ab}_{\al\gb}\Gamma^{cd}_{\gga\gd}\,,\quad
C_{ab,cd}=\ff{1}{64}C_{\al\gb\gga\gd}\Gamma^{\al\gb}_{ab}\Gamma^{\gga\gd}_{cd}\,.
\ee

To rewrite Cartan equations \eqref{Cartan1}-\eqref{Cartan2} in the
spinor form, we introduce the spinor symmetric Lorentz connection
one-form $\w_{\al\gb}=\Gamma_{\al\gb}^{ab}\w_{ab}$ and the
antisymmetric traceless f\"{u}nfbein one-form
$\e_{\al\gb}=\gga^{a}_{\al\gb}\e_a$. Equations
\eqref{Cartan1}-\eqref{Cartan2} then read
\begin{eqnarray}
\R_{\al\gb} &=&
d\w_{\al\gb}+\ff14\w_{\al}{}^{\gga}\wedge\w_{\gga\gb}\,,\\
\D\e_{\al\gb} &=&
d\e_{\al\gb}+\ff14\w_{\al}{}^{\gga}\wedge\e_{\gga\gb}+
\ff14\w_{\gb}{}^{\gga}\wedge\e_{\al\gga}=0\,,
\end{eqnarray}
where $\R_{\al\gb}=\R_{ab}\Gamma^{ab}_{\al\gb}\,$ and
$\D\xi_{\al}=d\xi_{\al}+\ff14\w_{\al}{}^{\gb}\xi_{\gb}$. The
integrability condition \eqref{int} reduces to
\be
\D^2\xi_{\al}=\ff14\R_{\al}{}^{\gb}\xi_{\gb}\,,\quad
\R_{\al}{}^{\gga}\wedge\e_{\gga\gb}-\R_{\gb}{}^{\gga}\wedge\e_{\gga\al}=0\,,
\ee
while \eqref{Weyl} and \eqref{int2} to, correspondingly,
\be\label{weyl}
\R_{\al\gb}=\Lambda\E2_{\al\gb}+\ff{1}{16}C_{\al\gb\gga\gd}\E2^{\gga\gd}\,,
\ee
\be
\D^2\xi_{\al}=\ff14\Lambda\E2_{\al}{}^{\gb}\xi_{\gb}-\ff{1}{64}
C_{\al\gb\gga\gd}\xi^{\gb}\E2^{\gga\gd}\,,
\ee
where
$\E2_{\al\gb}=\E2_{\gb\al}=\e_{\al}{}^{\gga}\wedge\e_{\gga\gb}$.
The following property resulting from \eqref{prop} has been used
\be
\e_{\al\gb}\wedge\e_{\gga\gd}=\ff12(\gep_{\al\gga}\E2_{\gb\gd}
-\gep_{\gb\gga}\E2_{\al\gd}-\gep_{\al\gd}\E2_{\gb\gga}
+\gep_{\gb\gd}\E2_{\al\gga})\,.
\ee

In establishing the spinor representation for Lorentz tensors one
has to force the reality condition which has not been considered
so far. Using that Hermitian conjugation for $\gamma$-matrices can
be expressed as
\be
\gga^{\dagger}_{a\,\al}{}^{\gb}=(\gga_0\gga_a\gga_0)_{\al}{}^{\gb}\,,\quad
(\Gamma^{\dagger}_{ab})_{\al}{}^{\gb}=(\gga_0\Gamma_{ab}\gga_0)_{\al}{}^{\gb}
\ee
we introduce the transformation
\be
T^D_{\al\gb}=(\gga_0T\gga_0)_{\al\gb}\,.
\ee
Now, $T_{\al\gb}=-T_{\gb\al}$ corresponds to the real vector $T^a$
given it is traceless and
\be\label{vec-real}
T^{\dagger}_{\al\gb}=T^D_{\al\gb}\,.
\ee
Analogously, $F_{\al\gb}=F_{\gb\al}$ is equivalent to the real
antisymmetric tensor $F_{ab}=-F_{ba}$ if
\be
F^{\dagger}_{\al\gb}=F^{D}_{\al\gb}\,.
\ee
Note, that a three-form $B_{abc}=B_{[abc]}$ has a symmetric
counterpart $B_{\al\gb}=B_{\gb\al}$; however, the reality
condition is different, namely
$B^{\dagger}_{\al\gb}=-B^{D}_{\al\gb}$. This means, in particular,
that the Hodge dualization $*F_{abc}=\varepsilon_{abcde}F^{de}$ is
reached by $*F_{\al\gb}=iF_{\al\gb}$ in the spinor notation.

\renewcommand{\theequation}{C.\arabic{equation}}
\renewcommand{\thesubsection}{C}
\renewcommand{\thesection}{Appendix}
\makeatletter \@addtoreset{equation}{subsection} \makeatother

\subsection{Proof of the geodesity condition}\label{App-C}
Let us prove \eqref{geodetic} for the Kerr-Schild vector $k^a$
\be\label{k-geo}
k^aD_ak_b=0\,.
\ee
The proof for $n^a$ is analogous. Before we start, one comment is
in order. In what follows, we only need  \eqref{main2} of the two
main equations \eqref{main1}-\eqref{main2}. We also require $v^a$
to be a Killing vector. This means that the geodesity condition
will be valid for any covariant derivatives in \eqref{k-geo}
either $D$ or $\D$.

The main idea for proving \eqref{k-geo} is first to prove that the
vector $t^a=k^bD_bk^a$ is light-like, {\it i.e.,}
\be\label{t2}
t^at_a=0\,.
\ee
In Lorentz signature, two orthogonal light-like vectors are
proportional. Since, by definition, $t_ak^a=0$ it implies
\be\label{t}
k^bD_bk_a\sim k_a,.
\ee
The unknown factor can be found by converting \eqref{t} with
$v^a$. Using that $D_av_b+D_bv_a=0$ and $v^ak_a=const=1$ the
factor is fixed to be zero.

To prove \eqref{t2} we need an auxiliary lemma:
\be
\Phi_{ab}k^b\sim k_a
\ee
or in the spinor form
\be\label{fk}
s_{\al\gb}\equiv\Phi_{\al}{}^{\gga}k_{\gga\gb}-\Phi_{\gb}{}^{\gga}k_{\gga\al}=Ak_{\al\gb}\,,
\ee
where $A$ is some factor. Since $s_{\al\gb}k^{\al\gb}=0$, it
suffices to prove that $s_{\al\gb}s^{\al\gb}=0$. From \eqref{KS}
it follows $\Pi^{-}_{\al}{}^{\gga}k_{\gga\gb}=0$ or
\be\label{Xk}
X_{\al}{}^{\gga}k_{\gga\gb}=k_{\al\gb}\,,
\ee
where $X_{\al\gb}$ was defined in \eqref{X}. Multiplying
\eqref{fk} by $X_{\gd}{}^{\al}$ we have
\be\label{Xs}
X_{\gd}{}^{\al}s_{\al\gb}=X_{\gd}{}^{\al}\Phi_{\al}{}^{\gga}k_{\gga\gb}-
\Phi_{\gb}{}^{\gga}X_{\gd}{}^{\al}k_{\gga\al}\,.
\ee
Noting that
$X_{\gd}{}^{\al}\Phi_{\al\gga}=-X_{\gga}{}^{\al}\Phi_{\al\gd}$
because $(X\Phi)_{\al\gb}$ contains only even powers of $\Phi$
which are antisymmetric and making use of \eqref{Xk}, equation
\eqref{Xs} amounts to
\be
X_{\al}{}^{\gga}s_{\gga\gb}=s_{\al\gb}\,.
\ee
Squaring it, one immediately obtains $s_{\al\gb}s^{\al\gb}=0$ that
proves \eqref{fk}.

Now we are in a position to prove the geodesity condition
\eqref{k-geo}. This results from the following relations
\begin{eqnarray}\label{proof}
\Phi_{\al}{}^{\gga}t_{\gga\gb}-\Phi_{\gb}{}^{\gga}t_{\gga\al}=
\ff14(\Phi_{\al}{}^{\gga}k^{\mu\nu}D_{\mu\nu}k_{\gga\gb}-
(\al\leftrightarrow\gb))=\nn\\
=\ff14\Big(k^{\mu\nu}D_{\mu\nu}(\Phi_{\al}{}^{\gga}k_{\gga\gb})
-k^{\mu\nu}k_{\gga\gb}D_{\mu\nu}\Phi_{\al}{}^{\gga}-(\al\leftrightarrow\gb)
\Big)\stackrel{\eqref{fk},\, \eqref{main2}}{=}\\
=\ff14k^{\mu\nu}D_{\mu\nu}(Ak_{\al\gb})+k_{\al}{}^{\gga}k_{\gb}{}^{\gd}
v_{\gga\gd}=At_{\al\gb}+k_{\al\gb}k^aD_aA+
k_{\al}{}^{\gga}k_{\gb}{}^{\gd}v_{\gga\gd}\nn
\end{eqnarray}
The bispinor
$u_{\al\gb}=k_{\al}{}^{\gga}k_{\gb}{}^{\gd}v_{\gga\gd}$ is
traceless, antisymmetric and yet satisfy the reality condition
\eqref{vec-real}, thus, being a vector. It is orthogonal to
$k_{\al\gb}$ and is light-like; hence $u_{\al\gb}\sim k_{\al\gb}$.
As a result \eqref{proof} amounts to
\be\label{end}
\Phi_{\al}{}^{\gga}t_{\gga\gb}-\Phi_{\gb}{}^{\gga}t_{\gga\al}=At_{\al\gb}
+Bk_{\al\gb}\,,
\ee
where $B$ is some irrelevant coefficient. From \eqref{end} it
follows $t_{\al\gb}t^{\al\gb}=0$ thus concluding the proof of
\eqref{k-geo}. Let us stress that the presented proof relies on
the Lorentz signature. Therefore, the spinor reality condition
plays an important role. We have omitted its explicit check here.

\renewcommand{\theequation}{D.\arabic{equation}}
\renewcommand{\thesubsection}{D}
\renewcommand{\thesection}{Appendix}
\makeatletter \@addtoreset{equation}{subsection} \makeatother

\subsection{Useful identities} \label{App-D}

Many useful relations that have been used throughout the scope of
the paper are mere consequences of Fierz identities. Namely,
antisymmetrization over four spinorial indices is proportional to
Levi-Cevita symbol $\varepsilon_{\al\gb\gga\gd}$ being expressed
as the antisymmetrization of the product of two charge-conjugation
matrices
\be
\varepsilon_{\al\gb\gga\gd}\sim(\gep_{\al\gb}\gep_{\gga\gd}-\gep_{\gga\gb}\gep_{\al\gd}-
\gep_{\gd\gb}\gep_{\gga\al})
\ee
Particularly, for a vector $v_{\al\gb}$ one easily finds
\begin{eqnarray}
v_{\al\gb}v_{\gga\gd}-v_{\gga\gb}v_{\al\gd}-v_{\gd\gb}v_{\gga\al}
&=& -v^2 (\gep_{\al\gb}\gep_{\gga\gd}-\gep_{\gga\gb}\gep_{\al\gd}-
\gep_{\gd\gb}\gep_{\gga\al})\,,\\
v_{\al}{}^{\gb}v_{\gb\gga}&=&v^2\gep_{\al\gga}\,,\quad
v^2=\ff14v_{\al\gb}v^{\al\gb}\,.
\end{eqnarray}
Another useful formula that can be obtained in this fashion
relates $\Phi^{3}_{\al\gb}$ to $\Phi_{\al\gb}$ and
$\Phi^{-1}_{\al\gb}$
\be\Phi^3_{\al\gb}+2Q\Phi_{\al\gb}+H^2\Phi^{-1}_{\al\gb}=0\,.
\ee
It also allows us to express the scalar
$\Phi^{-1}_{\al\gb}\Phi^{-1\,\al\gb}$ via $H$ and $Q$ introduced
in \eqref{scal}
\be\label{D5}
\ff14\Phi^{-1}_{\al\gb}(\Phi^{-1})^{\al\gb}=\ff{Q}{H^2}\,.
\ee
Here we put the other identities of the sort we have used:
\begin{eqnarray}
\ff12(\Phi^2_{\al}{}^{\gga}k_{\gga\gb}-\Phi^2_{\gb}{}^{\gga}k_{\gga\al})=-Qk_{\al\gb}\,,\label{D6}\\
k^{a}D_{a}\ff1H=\ff{2r}{H^2}\,,\quad
D_{a}\Big(\ff{r k^a}{H^3}\Big)=-\ff{2Q}{H^4}\,,\\
F_{ab}k^b=\ff{r}{2H^2}k_a
\,,\quad d\Big(\ff{k_{\al\gb}}{H}\e^{\al\gb}\Big)=-F_{\al\gb}\E2^{\al\gb}\,,\\
Qv_{\al\gb}+\ff12(\Phi^2_{\al}{}^{\gga}v_{\gga\gb}-\Phi^2_{\gb}{}^{\gga}v_{\gga\al})
=-iI_2\gep_{\al\gb}\,,\label{D9}\\
Qv_{\al\gb}+\ff12H^2(\Phi^{-2}_{\al}{}^{\gga}v_{\gga\gb}-\Phi^{-2}_{\gb}{}^{\gga}v_{\gga\al})
=iI_2\gep_{\al\gb}\,,\\
H^2\Phi_{\al\al}\Phi^{-1}_{\gb\gb}v^{\al\gb}v^{\al\gb}=-4(I^{2}_{2}+QX)\,,\quad
H^2\Phi^{-1}_{\al\al}\Phi^{-1}_{\gb\gb}v^{\al\gb}v^{\al\gb}=4X \label{D11}\\
\e^a\wedge\e^b=\ff18\Gamma^{ab}_{\al\gb}\E2^{\al\gb}\,,\quad
\E2_{\al\gb}=\Gamma^{ab}_{\al\gb}\e_a\wedge\e_b\,.
\end{eqnarray}

\end{document}